\documentclass[english,onecolumn]{IEEEtran}
\usepackage[T1]{fontenc}
\usepackage[latin9]{inputenc}
\usepackage{color}
\usepackage{amsthm}
\usepackage{amsmath}
\usepackage{amssymb}
\usepackage{graphicx}

\makeatletter

\theoremstyle{plain}
\newtheorem{thm}{\protect\theoremname}
\theoremstyle{definition}
\newtheorem{defn}{\protect\definitionname}
\theoremstyle{remark}
\newtheorem{rem}{\protect\remarkname}
\theoremstyle{plain}
\newtheorem{cor}{\protect\corollaryname}
\theoremstyle{plain}
\newtheorem{prop}{\protect\propositionname}
\theoremstyle{plain}
\newtheorem{lem}{\protect\lemmaname}

\usepackage{xspace}
\usepackage{bbm}
\input{mathlig}

\newcommand{\muspace}{\mspace{1mu}}

\DeclareRobustCommand{\scond}{\mathchoice{\muspace\vert\muspace}{\vert}{\vert}{\vert}}
\mathlig{|}{\scond}

\DeclareRobustCommand{\discint}{\mathchoice{\mspace{-1.5mu}:\mspace{-1.5mu}}{\mspace{-1.5mu}:\mspace{-1.5mu}}{:}{:}}
\mathlig{::}{\discint}

\newcommand{\Ac}{\mathcal{A}}

\newcommand{\Cc}{\mathcal{C}}

\newcommand{\Kc}{\mathcal{K}}

\newcommand{\Mc}{\mathcal{M}}

\newcommand{\Xc}{\mathcal{X}}
\newcommand{\Yc}{\mathcal{Y}}

\def\e{\epsilon}

\DeclareMathOperator\E{\textsf{E}}
\let\P\relax
\DeclareMathOperator\P{\textsf{P}}
\newcommand{\Bern}{\mathrm{Bern}}

\newcommand{\U}{\mathrm{Unif}}

\def\textiid{i.i.d.\@\xspace}
\newcommand\iid{\ifmmode\text{ i.i.d. } \else \textiid \fi}

\def\mathllap{\mathpalette\mathllapinternal}
\def\mathllapinternal#1#2{%
  \llap{$\mathsurround=0pt#1{#2}$}}

\def\clap#1{\hbox to 0pt{\hss#1\hss}}
\def\mathclap{\mathpalette\mathclapinternal}
\def\mathclapinternal#1#2{%
  \clap{$\mathsurround=0pt#1{#2}$}}

\let\oldstackrel\stackrel
\renewcommand{\stackrel}[2]{\oldstackrel{\mathclap{#1}}{#2}}

\renewcommand{\hbar}{h\mathllap{\overline{\vphantom{h}\hphantom{\rule{4.6pt}{0pt}}}\mspace{0.77mu}}}

\catcode`~=11 % make LaTeX treat tilde (~) like a normal character
\newcommand{\urltilde}{\kern -.06em\lower -.06em\hbox{~}\kern .02em}
\catcode`~=13 % revert back to treating tilde (~) as an active character

\usepackage{babel}
\providecommand{\corollaryname}{Corollary}
\providecommand{\definitionname}{Definition}
\providecommand{\lemmaname}{Lemma}
\providecommand{\propositionname}{Proposition}
\providecommand{\remarkname}{Remark}
\providecommand{\theoremname}{Theorem}

\makeatother

\usepackage{babel}
\providecommand{\corollaryname}{Corollary}
\providecommand{\definitionname}{Definition}
\providecommand{\lemmaname}{Lemma}
\providecommand{\propositionname}{Proposition}
\providecommand{\remarkname}{Remark}
\providecommand{\theoremname}{Theorem}

\begin{document}

\title{Maximal Correlation Secrecy}

% \author{Cheuk Ting Li, {\it Student Member, IEEE}, and Abbas El Gamal, {\it Fellow, IEEE}\\
% \thanks{The work of C. T. Li was supported in part by a Hong
% Kong Alumni Stanford Graduate Fellowship. This paper was presented in part at the IEEE International
% Symposium on Information Theory, Hong Kong, June 2015.%

% The authors are with the Department of Electrical Engineering, Stanford University, Stanford, CA 94305, USA (e-mail: \mbox{ctli@stanford.edu}, \mbox{abbas@ee.stanford.edu}).
% }
% Department of Electrical Engineering\\
% Stanford University\\
% Stanford, California, USA\\
% Email: ctli@stanford.edu, abbas@ee.stanford.edu
% }

\author{Cheuk Ting Li and Abbas El Gamal\\
 Department of Electrical Engineering\\
 Stanford University\\
 Stanford, California, USA\\
 Email: ctli@stanford.edu, abbas@ee.stanford.edu
 \thanks{The work of C. T. Li was supported in part by a Hong
 Kong Alumni Stanford Graduate Fellowship. This paper was presented in part at the IEEE International
 Symposium on Information Theory, Hong Kong, June 2015.}}

\maketitle
\begin{abstract}
This paper shows that the Hirschfeld-Gebelein-Rényi maximal correlation between the message and the ciphertext provides good secrecy guarantees for cryptosystems that use short keys. We first establish a bound on the eavesdropper's advantage in guessing functions of the message in terms of maximal correlation and the Rényi entropy of the message. This result implies that maximal correlation is stronger than the notion of entropic security introduced by Russell and Wang. We then show that a small maximal correlation $\rho$ can be achieved via a randomly generated cipher with key length $\approx2\log(1/\rho)$, independent of the message length, and by a stream cipher with key length $2\log(1/\rho)+\log n+2$ for a message of length $n$. We establish a converse showing that these ciphers are close to optimal. This is in contrast to entropic security for which there is a gap between the lower and upper bounds. Finally, we show that a small maximal correlation implies secrecy with respect to several mutual information based criteria but is not necessarily implied by them. Hence, maximal correlation is a stronger and more practically relevant measure of secrecy than mutual information. 
\end{abstract}

\section{Introduction\label{sec:Introduction}}

Consider the symmetric-key cryptosystem setting in which Alice encrypts
a message (plaintext) $M$ of length $n$ bits using a shared secret
key $K$ of length $s$ bits into a ciphertext $C$ and sends it to
Bob who recovers the message using the ciphertext and the key. The
system is said to provide perfect secrecy if the eavesdropper Eve
cannot gain {\em any} information about the message from the ciphertext
$C$ alone, that is, if $M$ and $C$ are independent. Shannon~\cite{Shannon1949b}
showed that achieving perfect secrecy requires the key to be as long
as the message. This result is considered impractical for most cryptographic applications and much shorter keys than the message are commonly used. 

To analyze the secrecy of cryptosystems that use short keys, less stringent criteria than perfect secrecy have been proposed. One such criterion is to limit Eve's ability to guess functions of $M$, by requiring that the difference between  Eve's probability of correctly guessing a function $f(M)$ of the message by a function of the ciphertext $\tilde{f}(C)$ and the maximum probability of correctly guessing $f(M)$ without knowledge of $C$, referred to as the \emph{advantage} of Eve
\begin{align}\label{eq:adv}
\mathrm{Adv}(f,\tilde f)=\P\left\{ f(M)=\tilde{f}(C)\right\} -\max_{i}\P\left\{ f(M)=i\right\}
\end{align}
to be small. 
While perfect secrecy is equivalent to requiring the advantage to be less than or equal to zero for all functions of $M$, we show that requiring the advantage to be less than a small positive value for all functions can be satisfied by keys that are much shorter than the message. In semantic security~\cite{goldwasser1984probabilistic}, a small advantage is required with the additional restriction that Eve uses only probabilistic polynomial-time algorithms. Although satisfied by short keys, proofs
of semantic security rely on unproven computational hardness assumptions. The closest work to this paper is entropic security introduced by Russell and Wang~\cite{russell2002fool} and studied
by Dodis and Smith \cite{dodis2005entropic}, which requires a small advantage
assuming the min-entropy of $M$ is large. They proposed several ciphers
with short keys that achieve entropic security and established lower bounds on the key length needed to achieve
entropic security. Their lower bounds and achievability results do not match, however (refer to Remark \ref{rmk:ent2} in Section~\ref{subsec:Definitions-and-Summary-Keylen} for details).

%We show that maximal correlation secrecy is stronger than these measures.} 

In this paper, we show that the Hirschfeld-Gebelein-Rényi
maximal correlation~\cite{hirschfeld1935connection,gebelein1941statistische,renyi1959measures}
between the message and the ciphertext, defined as 
\begin{equation}
\rho_{\mathrm{m}}(M;C)=\max_{\substack{f(m),g(c):\,\E(f(M))=\E(g(C))=0,\\
\E(f^{2}(M))=\E(g^{2}(C))=1
}
}\E\bigl(f(M)g(C)\bigr),\label{eq:mc}
\end{equation}
is a natural measure of secrecy for ciphers with short keys. We say
that a cipher achieves \emph{$\rho$-maximal correlation
secrecy} if $\rho_{\mathrm{m}}(M;\, C)\le\rho$ when $M$ is uniformly distributed.

Ciphers achieving maximal correlation secrecy can
guarantee a small advantage. If $M$ is uniformly distributed and
$f$ is a one-bit function, e.g., one of the bits of the message,
then the relationship between maximal correlation and the advantage
follows readily by the work of Witsenhausen~\cite{Witsenhausen1975}. Applying
the result by Calmon \textit{et al.}~\cite{Calmon2013inference},
the advantage for uniformly distributed $M$ and general $f$ is upper-bounded by $\rho$. A contribution of this paper is to extend this result on the relationship between maximal correlation and the advantage to scenarios in which the distribution of $M$ is not fixed and Eve has access to some side information about the message.
%\textcolor{blue}{The role of maximal correlation and principal inertia components in security and privacy was also studied in~\cite{Calmon2013inference,Calmon2014inertia,calmon2015fundamental}.}

Maximal correlation has numerous applications in information theory and statistics; see~\cite{Anantharam2013maxcor} for an overview. The work of Zhao and Chia~\cite{zhao2011} relates maximal correlation and secret key generation. The role of maximal correlation and principal inertia components in security and privacy was investigated by Calmon \textit{et al.}~\cite{Calmon2013inference,Calmon2014inertia,calmon2015fundamental}.

%More importantly, we show that maximal correlation $\rho$
%can be achieved using a key length $s\approx2\log(1/\rho)$ independent
%of the message length. Combining these two results, we can guarantee
%a very small advantage for Eve using a short key. 

Several other information theoretic secrecy criteria that do not require long keys have also been proposed.
%Several information theoretic measures of secrecy for cryptosystems with short keys have also been proposed.
%The notions of strong secrecy~\cite{Maurer94thestrong}, weak secrecy~\cite{Wyner1975d}, and leakage rate~\cite{Csiszar--Korner1978} use the mutual
%information between the message and the ciphertext. 
%\cite{csiszar1996almost,maurer2000information}
In~\cite{Wyner1975d}, Wyner proposed the wiretap channel where Eve observes a noisy version of the ciphertext.
In~\cite{Ozarow--Wyner1985}, Ozarow and Wyner studied the wiretap channel setting in which Eve can choose a subset of the ciphertext bits to observe.
In such settings, secrecy criteria based on the mutual information between the message and the eavesdropper's observation (e.g., weak secrecy and strong secrecy~\cite{Maurer94thestrong,csiszar1996almost,maurer2000information}) are typically used.
Semantic security has also been applied to wiretap channels~\cite{bellare2012semantic} without limitations on Eve's computational power.
In~\cite{massey1985rip}, Massey and Ingemarsson proposed a cipher with a long transmission delay which is secure assuming a memory constraint on Eve. In~\cite{cachin1997unconditional}, Cachin and Maurer proposed a cipher assuming a memory constraint on Eve, but without long delay.
%%In~\cite{maurer1991local}, Maurer and Massey ....
%%(which imposes a restriction that the eavesdropper can only observe a small number of ciphertext bits).
In~\cite{maurer1991perfect}, Maurer considered the scenario where Alice, Bob and Eve observe a random source over different noisy channels. In~\cite{maurer1992conditionally}, Maurer studied the case where there is a large public random source, and the number of bits in the random source that Eve can examine is limited. In~\cite{Calmon2012lists,calmon2015hiding}, Calmon~\textit{et al.} proposed a secrecy criterion called $\epsilon$-symbol secrecy which limits Eve's knowledge on subsets of bits of the message.
Note that the security criteria in the above works either depend on the bit structure of the message or ciphertext (e.g., noise is applied to the bits in wiretap channel, and $\epsilon$-symbol secrecy aims at protecting bits of the message), or impose a memory constraint. In contrast, $\rho$-maximal correlation secrecy (like entropic security) does not depend on the bit structure of the message, guaranteeing that the cipher hides every function of the messages (not only bits) equally well from an eavesdropper with unlimited memory and computational power.

\begin{figure}[h]
\begin{center}
\includegraphics{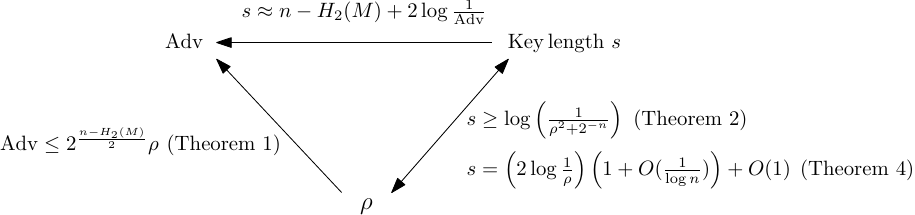}
\par\end{center}
\caption{\label{fig:relation}Summary of the relationship between $\mathrm{Adv}$, $s$ and $\rho$, given in Theorem~\ref{thm:binary_guess}, \ref{thm:converse} and \ref{thm:rand_achieve}.}
\end{figure}

The main contributions of this paper, which is an extended and more complete version of~\cite{mcsecurity_isit}, are as follows (also see Figure~\ref{fig:relation}).

\smallskip{}

\noindent \emph{Rényi entropy constrained security}.
In Section~\ref{subsec:Definitions-and-Summary-Security},
we define the notion of Rényi entropy constrained security and show that a $\rho$-maximal correlation secure cipher also achieves a variant of semantic security with computationally unbounded adversary.
In Theorem~\ref{thm:binary_guess}, we show that for non-uniform $M$, the advantage
is upper-bounded by $2^{(n-H_{2}(M))/2}\rho$,
where $H_{2}(M)$ is the Rényi entropy of $M$. Therefore we are
able to provide secrecy guarantees for a cipher used on data with
different pmfs, and even if some partial information about $M$ is
provided to Eve.
We further show that a small maximal correlation secrecy
implies entropic security (refer to Remark~\ref{rmk:ent1} in Section~\ref{subsec:Definitions-and-Summary-Security}
for details).
%In Section~\ref{subsec:Definitions-and-Summary-Security},
%we formally define the Rényi entropy constrained security setting.
The proof of our result is given in Section~\ref{subsec:Proofs-Thm-Binaryguess}.

\noindent \smallskip{}

\noindent \emph{Maximal correlation secrecy key length}. In Section~\ref{subsec:Definitions-and-Summary-Keylen},
we show the surprising result that $\rho$-maximal correlation secrecy
can be achieved by short keys of length independent of the message
length. We first establish a converse result showing that every $\rho$-maximal
correlation secure cipher must have a key length $s \ge 2\log(1/\rho)-\log\left(1+2^{-n}\rho^{-2}\right)$
bits (Theorem \ref{thm:converse}). We then show that a cipher constructed using expander graphs
can achieve $\rho$-maximal correlation secrecy with a key length
\mbox{$s=(2+O(1/ \log n))\log(1/\rho) + O(1)$} as $n\to \infty$ (Theorem \ref{thm:rand_achieve}). We
further show that $\rho$-maximal correlation secrecy can be achieved
with high probability via a randomly generated binary additive stream
cipher with a key length $s=2\log(1/\rho)+\log n+2$ (Theorem \ref{thm:achieve_binaryadd}).
%These results make maximal correlation secrecy relevant to practical cryptosystems,
%which typically use keys of fixed length. ????relationship to entropic security bounds here???
These results show that the tradeoff $s\approx 2\log(1/\rho)$ is optimal for large $n$. In contrast, the lower bounds on the key length for entropic security is not close to the achievable key length (refer to Remark~\ref{rmk:ent2} in Section~\ref{subsec:Definitions-and-Summary-Keylen} for details).
The proofs of these results are given in Section~\ref{sec:Proofs-of-Theorems}. 
For example, for a 1GB message and a 512-bits key, an advantage can be bounded by $\approx10^{-70}$.

\noindent \smallskip{}

\noindent \emph{Relationship to other secrecy criteria}. In Section~\ref{subsec:Definitions-and-Summary-Relationship}
we show that $\rho$-maximal correlation secrecy is a stronger measure of secrecy than the notions of strong secrecy~\cite{Maurer94thestrong,maurer2000information}, weak secrecy~\cite{Wyner1975d}, and leakage rate~\cite{Csiszar--Korner1978}, which use the mutual information between the message and the ciphertext. We show that these measures are implied by $\rho$-maximal correlation secrecy with suitable choices of $\rho$, but do not imply $\rho$-maximal
correlation secrecy for any $\rho<1$. \smallskip{}

%?????

% \subsection{Related Work}
% %
% %The maximal correlation was investigated by Hirschfeld~\cite{hirschfeld1935connection}, Gebelein~\cite{gebelein1941statistische} and R\'enyi~\cite{renyi1959measures}. Anantharam \textit{et al.}~\cite{Anantharam2013maxcor} gave an overview of different characterizations and properties of maximal correlation.
% %
% 
% Maximal correlation has numerous applications in information theory and statistics. Refer to Anantharam \textit{et al.}~\cite{Anantharam2013maxcor} for an overview. The work of Zhao and Chia~\cite{zhao2011} relates maximal correlation and secret key generation. The relationship between maximal correlation, principal inertias and function estimation was investigated by Calmon \textit{et al.}~\cite{Calmon2013inference,Calmon2014inertia}.
% 
% Other information-theoretic definitions of secrecy which allow the key to be shorter than the message were also proposed, for example, by Maurer~\cite{maurer1992conditionally}, Cachin and Maurer~\cite{cachin1997unconditional}, and Calmon \textit{et al.}~\cite{Calmon2012lists,Calmon2014infothmetric}.
% 

\section{Definitions and Notation}

%\noindent{\em Notation}.
Throughout this paper, we denote the joint probability matrix of $X$
and $Y$ by $P_{X,Y}\in\mathbb{R}^{\left|\Xc\right|\times\left|\Yc\right|}$.
We denote the spectral norm of the matrix $A\in\mathbb{R}^{m\times n}$
as 
\[
\left\Vert A\right\Vert =\max_{v\in\mathbb{R}^{n},\,\left\Vert v\right\Vert =1}\left\Vert Av\right\Vert 
\]
and its Frobenius norm as $\left\Vert A\right\Vert _{F}$. We denote
the $m\times n$ matrix consisting of all ones by $\mathbf{1}_{m\times n}$.
The $\log$ function is base 2 and the entropy is measured in bits.
We use the notation $[1:n]=\{1,2,\ldots,n\}$ and $\U(\Ac)$ to be
the uniform probability mass function (pmf) over a finite set $\Ac$.
%---------------------------------------

We consider a cryptosystem that consists of 
\begin{itemize}
\item  a message $M\in\Mc$, where $\Mc=[1:2^{n}]$, i.e., $M$ is an $n$-bit
message, unless specified otherwise, 
\item  a random secret key $K\sim\U(\Kc)$, where $\Kc=[1:2^{s}]$ unless
specified otherwise, 
\item  an encryption function $E(k,m)$ that maps every pair $(k,m)\in\Kc\times\Mc$
into a ciphertext $c\in\Cc$, where $\Cc=[1:2^{n}]$ unless specified
otherwise, and 
\item  a decryption function $D(k,c)$ that maps every pair $(k,c)\in\Kc\times\Cc$
into a message $m\in\Mc$ such that $D(k,E(k,m))=m$ for any $m,k$. 
\end{itemize}
The pair of encryption and decryption functions $(E,D)$ is called
a (block) cipher. We assume throughout the paper that the eavesdropper
knows the ciphertext $C$ but not the message $M$ or the key $K$.
A cipher is said to be $\rho$-maximal correlation secure if $\rho_{\mathrm{m}}(M,E(K,M))\le\rho$
assuming that $M\sim\U(\Mc)$, where $\rho_{\mathrm{m}}$ is as defined
in~\eqref{eq:mc}. The encryption function can also be probabilistic.
In this case, the ciphertext $C=E(K,M,W)$ is also a function of a
random variable $W$, which is generated using the local randomness
at the sender, and is unknown to the receiver and the eavesdropper.
The cipher is assumed to be deterministic unless specified otherwise.

%--------------

\section{Rényi Entropy Constrained Security\label{subsec:Definitions-and-Summary-Security}}
%WHY TALK ABOUT SEMANTIC SECURITY? Lets rewrite 
In this section we show that every $\rho$-maximal
correlation secure cipher satisfies a variant of semantic security.
Recall that a cipher is said to be semantically secure~\cite{goldreich2009foundations}
if for any pmf $p(m)$ on $M$, any function $f(m)$, and any partial
information function $h(m)$ of the message, if $M$ is generated
according to $p(m)$, the eavesdropper who observes the ciphertext
$C$ and $h(M)$ (and also knows the choices of $n$, $p$, $f$ and
$h$) cannot correctly guess $f(M)$ using a probabilistic, polynomial-time
algorithm with probability non-negligibly higher than the best probabilistic,
polynomial-time algorithm for guessing $f(M)$ using only $h(M)$
(and also the choices of $n$, $p$, $f$ and $h$). In other words,
the eavesdropper cannot improve the probability of guessing $f(M)$
correctly by observing $C$. Note that the definition in~\cite{goldreich2009foundations}
allows $p(m)$ to be a pmf on messages with different lengths. For
simplicity, we consider $p(m)$ to be a pmf on messages with the same
length $n$.

We assume $p(m)$ is a pmf on message with the same length $n$, and leave out the computational complexity assumptions on $p$, $f$ and $h$ here since they are not the main concern of this paper.

Rényi entropy constrained security is a variant of
semantic security in which we remove the limitation on computational
power but restrict the choice of the pmf $p(m)$ to have a large Rényi
entropy $H_{2}(M)$ (or equivalently a small $\chi^{2}$-divergence~\cite{Pearson1900}
from the uniform pmf), that is,
\[
H_{2}(M)=-\log\sum_{m}\left(p(m)\right)^{2}\ge t
\]
for some $t\ge0$. It is formally defined below.
\begin{defn}
A cipher is said to be $(t,\e)$-Rényi entropy constrained
secure if for any pmf $p(m)$ with $H_{2}(M)\ge t$, any function
$f(m)$ of the message, and any eavesdropper's guess $\tilde{f}(c)$
of $f(m)$, when the message $M$ is generated according to $p(m)$,
the advantage as defined in~\eqref{eq:adv}
is bounded as
$\mathrm{Adv}(f,\tilde f)\le \e$. %=\P\{f(M)=\tilde{f}(C)\}-\max_{i}\P\{f(M)=i\}\le \e.

\end{defn}
%????Note that the first term is the probability that
%the eavesdropper can guess $f(M)$ correctly with knowledge of $C$,
%and the second term is the probability that the eavesdropper can guess
%it correctly without $C$ by fixing the outcome of $\tilde{f}(C)$.
The case of partial information $h(m)$ will be addressed later.

We now show that maximal correlation secrecy implies
Rényi entropy constrained security.
\begin{thm}
\label{thm:binary_guess} A $\rho$-maximal correlation
secure cipher is $(t,\e)$-Rényi entropy constrained secure for any
$t\ge0$ and{
\[
\epsilon=2^{(n-t)/2}\rho.
\]
} Moreover, if the choice of $f(m)$ is restricted to one-bit functions
$f(m)\in\{0,1\}$, then the advantage is bounded by{
\[
\epsilon=2^{(n-t)/2-1}\rho.
\]
}
\end{thm}
The proof of this theorem is given in Section \ref{subsec:Proofs-Thm-Binaryguess}.
Note that $\rho$ is always measured 
assuming the pmf of the message is uniform, though the theorem shows that $\rho$ can
also be used to bound the advantage when $p(m)$ is non-uniform.
Also note that the value of $n-t$ corresponds to the deviation of $p(m)$ from the uniform distribution,
and can be very small when $p(m)$ is close to uniform.

\smallskip{}

The Rényi entropy
constrained security can be extended to scenarios in which partial information $h(m)$ is 
available to Eve. We restrict the choices of $p(m)$ and $h(m)$ to
satisfy the condition 
\[
\sum_{a}\P\left\{ h(M)=a\right\} 2^{-H_{2}(M\,|\, h(M)=a)/2}\le2^{-\tau/2},
\]
where $\tau\ge0$ is a constant and $H_{2}(M\,|\, h(M)=a)$ is the
Rényi entropy of the conditional pmf of $M$ given $h(M)=a$.
The eavesdropper's guess $\tilde{f}(c,h(m))$ can depend on $h(M)$,
and the advantage is now defined as 
\[
\P\{f(M)=\tilde{f}(C,h(M))\}-\E\left(\max_{i}\P\{f(M)=i\,|\, h(M)\}\right).
\]
where the second term is the probability of guessing $f(M)$ correctly
using the maximum a posteriori estimation of $f(M)$ given $h(M)$.
As a consequence of Theorem \ref{thm:binary_guess}, for a $\rho$-maximal
correlation secure cipher, the advantage is upper bounded by $2^{(n-\tau)/2}\rho$.

The value of $\rho$ directly corresponds to the eavesdropper advantage
and the correct probability of the eavesdropper's guess. For example,
if the message $M$ is uniformly distributed, (i.e., $t=n$), then
the eavesdropper cannot correctly guess any one-bit function such
that $\P\{f(M)=1\}=1/2$ with probability larger than $\left(1+\rho\right)/2$.
As another example, if $M$ is uniformly distributed and $l$ bits
of $M$ (at fixed positions) are provided to the eavesdropper via
the partial information $h(m)$, then the advantage of the eavesdropper
is upper bounded by $2^{l/2}\rho$.

To illustrate our results, suppose we wish to protect a message of
length $n=8\times10^{9}$ (i.e., 1GB) with a key of length $s=512$.
By Theorem \ref{thm:achieve_binaryadd}, we can achieve $\rho$-maximal
correlation secrecy for $\rho=1.54\times10^{-72}$ using a binary
additive stream cipher. As a result, if $M$ is uniformly distributed,
then the advantage of the eavesdropper is upper bounded by $1.54\times10^{-72}$.
If $l=100$ bits of $M$ are provided to the eavesdropper, then the
advantage is bounded by $1.74\times10^{-57}$. We can see that a cipher
with key length much shorter than the message length can provide good
security guarantees.
\begin{rem}
\label{rmk:ent1}

We can show that $\rho$-maximal correlation secrecy implies $(t,\e)$-entropic security (as defined in~\cite{dodis2005entropic}) for $\e = 2^{(n-t)/2} \rho$. In fact the implication holds even when the min-entropy $H_\infty (M)$ in entropic security is replaced by Rényi entropy $H_2(M)$, i.e., for any pmf $p(m)$ of $M$ with $H_{2}(M)\ge t$, and any function $\tilde{f}(c)$,
there exists a random variable $\tilde{F}$ independent of $M$ such that for any function
$f(m)$,
\[
\left| \P\{ f(M)=\tilde{f}(C)\} - \P\{ f(M)=\tilde{F}\} \right| \le \epsilon.
\]
Since $H_2(M) \ge H_\infty (M)$ and the difference can be quite large if one of the messages has a high probability, maximal correlation secrecy can be much stronger than entropic security. The proof is given in Section \ref{subsec:Proofs-Thm-Binaryguess}.
\end{rem}

\section{Maximal Correlation Secrecy Key Length\label{subsec:Definitions-and-Summary-Keylen}}

We provide bounds on the key length of a $\rho$-maximal correlation
secure cipher in terms of $\rho$ and the message length $n$. We
first establish the following lower bound on the key length.
\begin{thm}
\label{thm:converse} If a cipher is $\rho$-maximal correlation secure,
then its key length is lower bounded as{ %\[
%s \ge 2\log\frac{1}{\rho}-\log\left(1+\frac{1}{2^{n}\rho^{2}}\right).
%\]
\[
s\ge\log\left(\frac{1}{\rho^{2}+2^{-n}}\right).
\]
} 
\end{thm}
The proof of this theorem is given in Section~\ref{subsec:Proofs-Thm-Converse}.
Note that when $\rho>0$, this lower bound can be written as 
\[
s\ge2\log\frac{1}{\rho}-\log\left(1+\frac{1}{2^{n}\rho^{2}}\right),
\]
which approaches $2\log(1/\rho)$ as $n$ tends to infinity. Also
note that this bound applies to any ciphertext length (not necessarily
equal to message length) and to probabilistic encryption functions.

%?????The use of expander
%graphs and invertible extractors in constructing ciphers was also
%studied in \cite{dodis2005entropic,cheraghchi2012invertible}.

We now consider a construction of a cipher with key length close to the
lower bound using expander graphs similar to~\cite{dodis2005entropic,cheraghchi2012invertible}. Let
$G$ be a $d$-regular graph with vertices $[1:2^{n}]$ and edges
$(m,\, E(k,m))$ for $m\in[1:2^{n}]$, $k\in[1:d]$, where $E$ is
a labeling of the edges of $G$. The graph may be a multigraph with
multiple instances of the same edge, and we assume the graph is undirected,
i.e., the number of edges $(m,c)$ is the same as the number of edges
$(c,m)$. Further assume the labeling is invertible, i.e., there exists
function $D(k,c)$ such that $D(k,E(k,m))=m$ for all $m,k$. The
adjacency matrix of $G$ is given by
\[
A\in\mathbb{R}^{2^{n}\times2^{n}},\; A_{m,c}=\left|\left\{ k:\, E(k,m)=c\right\} \right|.
\]
Such a graph is referred to as an {\em expander graph} if the
magnitude of the second largest eigenvalue (in absolute value) of
$A$ 
\[
|\lambda_{2}(A)|=\left\Vert A-\frac{d}{2^{n}}\mathbf{1}_{2^{n}\times2^{n}}\right\Vert 
\]
is small. An expander graph can be constructed explicitly. For example,
a (non-bipartite) Ramanujan graph~\cite{lubotzky1988ramanujan} has
a second eigenvalue 
\[
|\lambda_{2}(A)|\le2\sqrt{d-1}.
\]
Given an expander graph, we can define a corresponding {\em expander
graph cipher} with $\Mc=\Cc=[1:2^{n}]$, $\Kc=[1:d]$, encryption
function $E(k,m)$, and decryption function $D(k,c)$. We now find
the maximal correlation for such an expander graph cipher. 
\begin{thm}
\label{thm:achieve_expander}The cipher defined by an expander graph
with adjacency matrix $A$ has maximal correlation{
\[
\rho_{\mathrm{m}}(M;C)=\frac{1}{d}|\lambda_{2}(A)|.
\]
} As a result, the cipher corresponding to a non-bipartite Ramanujan
graph is $\rho$-maximal correlation secure if{
\[
\log d\ge2\log\frac{1}{\rho}+2.
\]
} where $s=\log d$ corresponds to the key length if it is an integer. 
\end{thm}
\smallskip{}

The proof of this theorem is given in Section \ref{subsec:Proofs-Thm-Expander}.
It is a consequence of the characterization of maximal correlation
in~\cite{Witsenhausen1975}. The relationship between maximal correlation
and the second eigenvalue of a graph is also studied in~\cite{Bolla200423}.
%\begin{thm}
%\label{thm:achieve_expander} If the Cayley graph generated by $\Mc$ and $\Kc$ is a Ramanujan graph, then the expander graph cipher .... is $\rho$-maximal correlation
%secure if
%\[
%\log | \Kc | \ge 2\log\frac{1}{\rho} + 2.
%\]
%%\[
%%\rho = \frac{2\sqrt{|\Kc|-1}}{|\Kc|}.
%%\]
%\end{thm}
%NEED TO STRAIGHTEN OUT NOTATION OF MC AND KC
A limitation of this construction is that there may not be constructions
of Ramanujan graphs for a desired $n$ and $s$. Using the result
in~\cite{Friedman1991} on the second eigenvalue of random regular
graphs, we can show the existence of maximal correlation secure ciphers
with key lengths close to the lower bound for any large enough $n$
and $s$.

\smallskip{}

\begin{thm}
\label{thm:rand_achieve}There exists a $\rho$-maximal correlation
secure cipher with message length $n\ge2$ and key length $s\ge2$
if {
\[
s\ge\left(2\log\frac{1}{\rho}\right)\left(1+\frac{\alpha}{\log n}\right)+\alpha,
\]
} where $\alpha>0$ is a constant. 
\end{thm}
The following corollary provides a bound on $s$ which is independent
of $n$. 
\begin{cor}
\label{cor:rand_achieve_rhoonly} There exists a $\rho$-maximal correlation
secure cipher with message length $n\ge2$ and key length $s\ge2$
if {
\[
s\ge\left(2\log\frac{1}{\rho}\right)\left(1+\frac{3\alpha/2}{\log(\log(1/\rho)+1)}\right),
\]
} where $\alpha>0$ is a constant. 
\end{cor}
\smallskip{}

The proofs of Theorem~\ref{thm:rand_achieve} and Corollary~\ref{cor:rand_achieve_rhoonly}
are in Section~\ref{subsec:Proofs-Thm-Rand}. This corollary shows
that for any $\rho$, a key length which depends only on $\rho$ is
sufficient to achieve $\rho$-maximal correlation secrecy for any
message length. This is in a strong contrast to perfect secrecy, which
requires the key length to be at least the message length.

Maximal correlation secrecy can also be achieved by a simpler cipher
with a sightly longer key length. Consider a \emph{binary additive
stream cipher} $\Mc=\Cc=\{0,1\}^{n}$, $\Kc=\{0,1\}^{s}$, $E(k,m)=m\oplus g(k)$,
$D(k,c)=c\oplus g(k)$, where $g(k)=(g_{1}(k),g_{2}(k),\ldots,g_{n}(k))\in\{0,1\}^{n}$
is the {\em keystream generator} and $\oplus$ is component-wise
$\mathrm{mod}\,2$ addition. The following theorem shows that most
binary additive stream ciphers with slightly longer key than the lower
bound in Theorem~\ref{thm:converse} are $\rho$-maximal correlation
secure.
\begin{thm}
\label{thm:achieve_binaryadd} Let $G_{i}(k)$, $i\in[1:n]$, $k\in[1:2^{s}]$
be i.i.d. $\Bern(1/2)$ random keystream components. Let $\rho>0$,
$\e>0$, then \textnormal{
\[
\P\{\rho_{\mathrm{m}}(M;\, M\oplus G(K))\le\rho\}>1-\e,
\]
} where the randomness of $\rho_{\mathrm{m}}(M;M\oplus G(K))$ is
induced by the random keystream generator, if the key length{
\[
s\ge2\log\frac{1}{\rho}+\log n+\log\left(1+\frac{1}{n}\log\frac{1}{\e}\right)+2.
\]
} 
\end{thm}
The proof of this theorem is given in Section \ref{subsec:Proofs-Thm-Binaryadd}.
Substituting $\e=1$ in the theorem shows that there exists a binary
additive stream cipher that is $\rho$-maximal correlation secure
with a key length 
\[
s\ge2\log\frac{1}{\rho}+\log n+2.
\]
Hence for a constant $\rho>0$, a key size of around $\log n$ is
sufficient.

\begin{figure}[htpb]
%\psfrag{A}[t]{$\rho$}
%\psfrag{B}[b]{Key length}
%\psfrag{C}[c]{Lower bound on key length of Theorem 1}
%\psfrag{D}[c]{Achievable key length of Theorem 2}
%\psfrag{E}[c]{Achievable key length of Theorem 4}
% \includegraphics[scale=0.45]{mcgraph4.eps}

\begin{centering}
\includegraphics[scale=0.308]{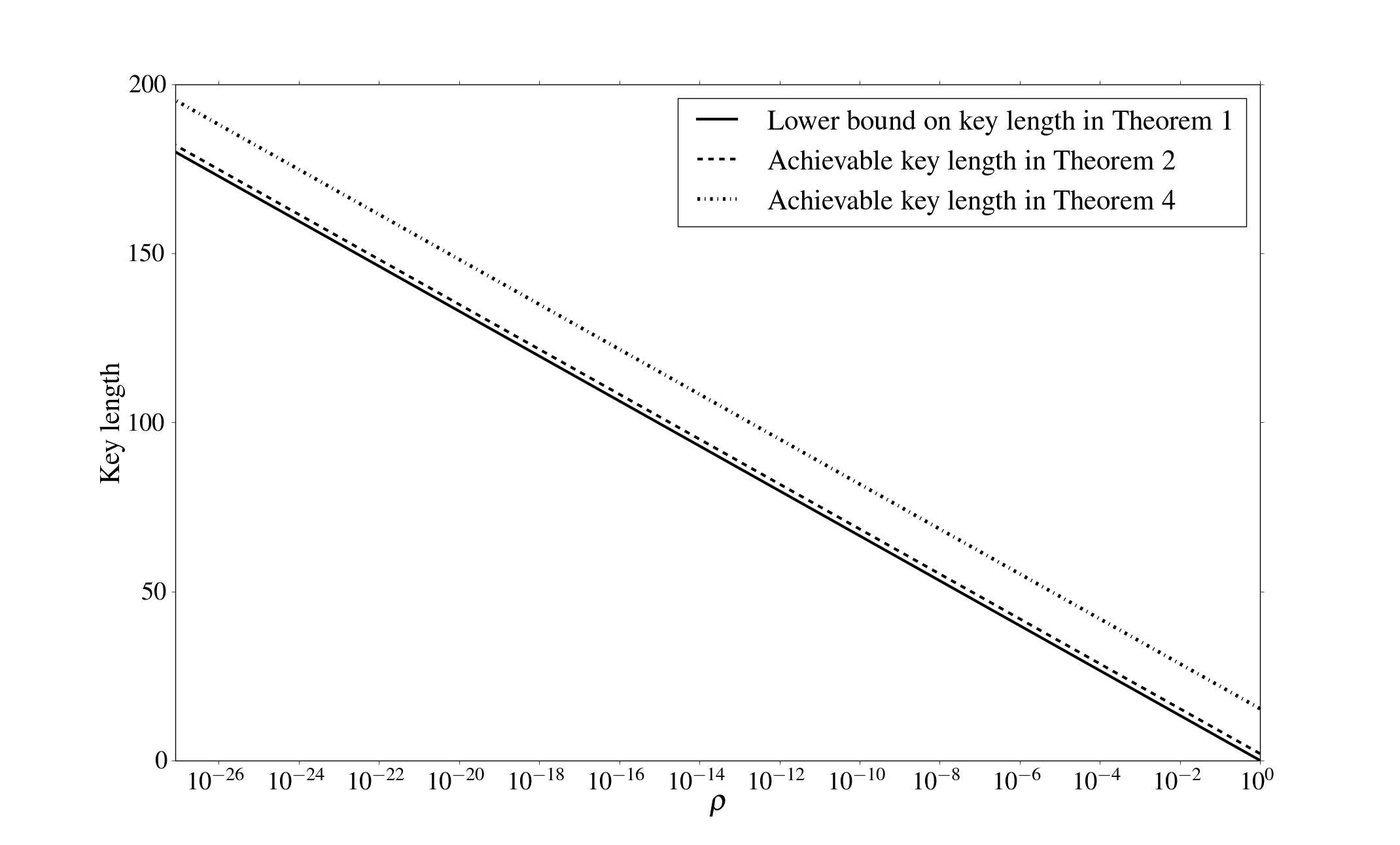} \caption{Comparison of the lower bound and the achievable key lengths for $n=10000$.}

\par\end{centering}

\centering{}\label{fig:s_compare} 
\end{figure}

Figure \ref{fig:s_compare} plots the lower bound on the key length
in Theorem \ref{thm:converse}, the key length achievable by the expander
graph cipher using the Ramanujan graphs in Theorem \ref{thm:achieve_expander},
and the key length achievable by the random stream cipher in Theorem
\ref{thm:achieve_binaryadd} versus $\rho$ for $n=10000$. 

%WE NEED TO RERWITE THIS REMARK 
\begin{rem}
\label{rmk:ent2}
%... ~\cite{dodis2005entropic,cheraghchi2012invertible} also used expander
%graphs and invertible extractors to construct ciphers ....

By the achievability result in
Theorem \ref{thm:achieve_expander} where $s=2\log(1/\rho)+2$, there
exist ciphers which achieve $(t,\e)$-Rényi entropy constrained security
with key length
\[
s=n-t+2\log\frac{1}{\e}+2.
\]
This is the same key length required by entropic
security in \cite{dodis2005entropic} (Corollary 3.3). Since
$(t,\e)$-Rényi
entropy constrained security implies $(t,\e)$-entropic security,
it is possible to show the achievability result of entropic security
via maximal correlation.

Also note that unlike maximal correlation
secrecy where there are tight upper and lower bounds on the optimal
key length $s\approx2\log(1/\rho)$ for large $n$, the lower bounds
on the key length given in \cite{dodis2005entropic}:
\[
s\ge n-t
\]
for any entropically secure cipher, and
\[
s\ge n-t+\log\frac{1}{\epsilon}-O(1)
\]
for public-coin schemes, are much smaller than the achievable key
length. It is conjectured in~\cite{dodis2005entropic} that the achievable key length is tight.
\end{rem}

\section{Relationship to other Secrecy Criteria\label{subsec:Definitions-and-Summary-Relationship}}

In this section, we compare maximal correlation secrecy
to other secrecy criteria that use the mutual information $I(M;C)$
between the message $M$ and the ciphertext $C$. We show that $\rho$-maximal
correlation secrecy is stronger than these other secrecy criteria
as they are implied by $\rho$-maximal correlation secrecy with suitable
choices of $\rho$, but they do not imply $\rho$-maximal correlation
secrecy for any $\rho<1$.

As pointed out by Maurer~\cite{Maurer94thestrong},
mutual information is too loose a criterion if it is not required
to approach zero because %The reason is that a cipher that achieves a desired mutual information is not guaranteed to keep any bit of the message, or more generally any function of the message, secure.
it does not guarantee hiding any  bit (or more generally any function) of the message. For an $n$ bit message and an $s$-bit key the mutual information is lower bounded as
$I(M;C)\ge n-s$ and this bound can be achieved simply by
encrypting the most significant $s$ bits of $M$ perfectly and sending
the rest of the bits in the clear. Hence unless $s\ge n$, mutual
information cannot guarantee that any bit of $M$ is not leaked.
There are other ciphers that can achieve the same minimum mutual information but ensure better secrecy. For example, suppose we wish to encrypt
a 2-bit message $M\sim\U(\{0,1,2,3\})$ using a 1-bit key $K\sim\Bern(1/2)$. Let $C_{1}=M+2K\mod4$ be the cipher which encrypts only the most
significant bit of the message and  
$C_{2}=M+K\mod4$ be another cipher that achieves $I(M;C)=1$.  It is easy to see that using $C_2$, Eve cannot correctly guess any bit of the message with probability greater than $3/4$ signifying that $C_{2}$ provides better secrecy than $C_{1}$. Moreover the second cipher achieves lower maximal correlation $\rho_{\mathrm{m}}(M;C_{2})=\sqrt{2}/2$
than  the first $\rho_{\mathrm{m}}(M;C_{1})=1$, hence maximal correlation is a better measure of secrecy than mutual information.

We now formalize the relationship between maximal correlation secrecy and the mutual information secrecy measures of strong secrecy~\cite{Maurer94thestrong},
weak secrecy~\cite{Wyner1975d}, and leakage rate~\cite{Csiszar--Korner1978}. By the following proposition which follows from Theorem 5 in \cite{gibbs2002choosing}, a $\rho$-maximal correlation secrecy guarantees
a small mutual information.
\begin{prop}
\label{prop:rho_mi_cipher} Let $X$ and $Y$ be two discrete random
variables, then {
\begin{equation}
I(X;Y)\le\log\left((\min\left\{ \left|\Xc\right|,\left|\Yc\right|\right\} -1)\cdot\rho_{\mathrm{m}}^{2}(X;Y)+1\right).\label{eq:mutualinfo-bound}
\end{equation}
}
\end{prop}
In the following we assume that $M\sim\U([1:2^{n}])$, which reduces~\eqref{eq:mutualinfo-bound}
to 
\begin{equation}
I(M;C)\le\log\left((2^{n}-1)\rho_{\mathrm{m}}^{2}(M;C)+1\right).\label{eq:mutualinfo-bound1}
\end{equation}
We now use the above proposition to compare $\rho$-maximal correlation
secrecy to secrecy criteria that use the mutual information. \smallskip{}

\noindent \emph{Strong secrecy}. This criterion requires that $\lim_{n\to\infty}I(M;C)=0$.
From~\eqref{eq:mutualinfo-bound1} this is implied by $\rho$-maximal
correlation secrecy for 
\[
\rho=o(2^{-n/2}).
\]
\smallskip{}

\noindent \emph{Weak secrecy}. This criterion requires that $\lim_{n\to\infty}I(M;C)/n=0$.
From~\eqref{eq:mutualinfo-bound1}, this is implied by $\rho$-maximal
correlation secrecy for 
\[
\rho=2^{-n/2+o(n)}.
\]
\smallskip{}

\noindent \emph{Leakage rate}. Note that both weak and strong secrecy
require the key rate $\lim_{n}s/n=1$. By requiring that $\lim_{n}I(M;C)/n\le R_{L}$
for some leakage rate $R_{L}$, a key rate of $1-R_{L}$ can be achieved.
From~\eqref{eq:mutualinfo-bound1}, this is implied by $\rho$-maximal
correlation secrecy for 
\[
\rho=2^{-(1-R_{L})n/2+o(n)}.
\]
Note that Theorem \ref{thm:achieve_binaryadd} implies that such $\rho$
can be achieved also by a key rate of $1-R_{L}$. Hence maximal correlation
secrecy provides a better security guarantee than leakage rate with
no penalty on the key rate. %\end{itemize}

\noindent \smallskip{}

The above results show that $\rho$-maximal correlation secret implies
secrecy criteria involving mutual information. We now show that a
small $I(M;C)$ does not necessarily imply $\rho$-maximal correlation
secrecy. Consider the following cipher: Let $\Mc=\Cc=\Kc=[0:2^{n}-1]$,
and the encryption and decryption functions be 
\[
E(k,m)=\begin{cases}
m+k\mod2^{n}-1 & \text{if }m<2^{n}-1\\
2^{n}-1 & \text{if }m=2^{n}-1,
\end{cases}
\]
and $D(k,c)=E(-k,c)$. Direct computation yields $I(M;C)=2^{-n}(n+2-2^{-(n-1)})$,
which goes to zero as $n$ tends to infinity, and thus the cipher
satisfies strong secrecy. However, since one can determine if $M=2^{n}-1$
or not by observing $C$, $\rho_{\mathrm{m}}(M;C)=1$. Hence $\rho$-maximal
correlation secrecy is a strictly stronger secrecy criterion than
criteria that use mutual information.

\section{Proof of the results \label{sec:Proofs-of-Theorems}}

\subsection{Properties of Maximal Correlation\label{subsec:Proofs-Properties}}

We use a characterization of maximal correlation in terms of
the spectral norm that follows directly from the singular value characterization
of maximal correlation in~\cite{Witsenhausen1975}.
\begin{lem}
\label{lem:rho_norm} Let $(X,Y)\sim p(x,y)$ be discrete random variables
with marginals $p(x)$ and $p(y)$. Define the matrix $B\in\mathbb{R}^{|\Xc|\times|\Yc|}$
with entries{
\[
B_{xy}=\frac{p(x,y)}{\sqrt{p(x)p(y)}}-\sqrt{p(x)p(y)}.
\]
} Then,{
\[
\rho_{\mathrm{m}}(X;Y)=\Vert B\Vert.
\]
} \end{lem}

Next we state a result relating maximal correlation
and the $\chi^{2}$-divergence between the joint pmf and the product
of the marginal pmfs, also known as $\chi^{2}$ measure of correlation which
follows directly from~\cite{Calmon2013inference}.
\begin{lem}
\label{lem:rho_chisq} Let $(X,Y)\sim p(x,y)$ be discrete random
variables with marginals $p(x)$ and $p(y)$. Then,{
\[
\frac{1}{\min\left\{ \left|\Xc\right|,\left|\Yc\right|\right\} -1}\le\frac{\rho_{\mathrm{m}}^{2}(X;Y)}{\chi^{2}\left(p(x,y)\,\Vert\, p(x)p(y)\right)}\le1,
\]
} where{
\[
\chi^{2}(p(x)\Vert q(x))=\sum_{\mathrm{m}}\frac{\left(p(x)\right)^{2}}{q(x)}-1
\]
} is the $\chi^{2}$-divergence between $p(x)$ and $q(x)$. \end{lem}

%----------------------------------------

\subsection{Proof of Theorem \ref{thm:binary_guess}\label{subsec:Proofs-Thm-Binaryguess}}

We prove Theorem \ref{thm:binary_guess}, which shows
that maximal correlation secrecy implies Rényi entropy constrained
security. First prove the case for general functions
$f(m)$. Note that it is implied by the following more general result. 
\begin{prop}
\label{prop:general_guess_dist} Consider any two
pmfs $p_{M}(m)$ and $\hat{p}_{M}(m)$ on $\Mc$, and a Markov kernel
$p_{C|M}(c|m)$. The two pmfs induce the joint probability measures
\textnormal{$\P$} and \textnormal{$\hat{\P}$} on $(M,C)$, respectively. Let $\rho_{\mathrm{m}}(M;C)$
be the maximal correlation in \textnormal{$\P$}. For any functions $f:\Mc\to\mathbb{N}$
and $\tilde{f}:\Cc\to\mathbb{N}$, we have \textnormal{
\[
\Big| \hat{\P}\left\{ f(M)=\tilde{f}(C)\right\} -\sum_{i} \hat{\P}\left\{ f(M)=i\right\} \cdot \P\{\tilde{f}(C)=i\} \Big| \le\rho_{\mathrm{m}}(M;C)\sqrt{\chi^{2}\left(\hat{p}_{M}\Vert p_{M}\right)+1}.
\]
} \end{prop}
\begin{IEEEproof}
All expectations, variances and covariances in this proof are in $\P$.
Assume the range of $f(m)$ is $\{1,...,l\}$. Let $Z_{1},...,Z_{l}$
be i.i.d. Rademacher random variables. Let $g(m)=Z_{f(m)}\hat{p}_{M}(m)/p_{M}(m)$
and $\tilde{g}(c)=Z_{\tilde{f}(c)}$. Write $\chi^{2}=\chi^{2}\left(\hat{p}_{M}\Vert p_{M}\right)$,
$\hat{p}_{f}(i)=\hat{\P}\left\{ f(M)=i\right\} $, $p_{\tilde{f}}(i)=\P\{\tilde{f}(C)=i\}$
and $\hat{p}_{eq}=\hat{\P}\{f(M)=\tilde{f}(C)\}$. Observe that 
\[
\E(g(M)\,|\, Z_{1}^{l})=\sum_{i}\hat{p}_{f}(i)Z_{i},
\]
\[
\E(\tilde{g}(C)\,|\, Z_{1}^{l})=\sum_{i}p_{\tilde{f}}(i)Z_{i},
\]
\begin{align*}
\E\left((g(M))^{2}\,|\, Z_{1}^{l}\right) & =\sum_{m}p_{M}(m)\left(\frac{Z_{f(m)}\hat{p}_{M}(m)}{p_{M}(m)}\right)^{2}\\
 & =\chi^{2}+1,
\end{align*}
\begin{align*}
\E\left(\mathrm{Cov}\big(g(M),\,\tilde{g}(C)\,|\, Z_{1}^{l}\big)\right) & =\E\left(\E\big(g(M)\tilde{g}(C)\,|\, Z_{1}^{l}\big)-\E(g(M)\,|\, Z_{1}^{l})\E(\tilde{g}(C)\,|\, Z_{1}^{l})\right)\\
 & =\E\left(\sum_{m}p(m)g(m)\sum_{c}p(c|m)\tilde{g}(c)-\big(\sum_{i}\hat{p}_{f}(i)Z_{i}\big)\big(\sum_{i}p_{\tilde{f}}(i)Z_{i}\big)\right)\\
 & =\E\left(\sum_{m}\hat{p}_{M}(m)\sum_{c}p(c|m)Z_{\tilde{f}(c)}Z_{f(m)}-\big(\sum_{i}\hat{p}_{f}(i)Z_{i}\big)\big(\sum_{i}p_{\tilde{f}}(i)Z_{i}\big)\right)\\
 & =\hat{p}_{eq}-\sum_{i}\hat{p}_{f}(i)p_{\tilde{f}}(i).
\end{align*}
\begin{align*}
\E\left( \left| \mathrm{Cov}\big(g(M),\,\tilde{g}(C)\,|\, Z_{1}^{l}\big)\right| \right) & \ge \left| \E\left(\mathrm{Cov}\big(g(M),\,\tilde{g}(C)\,|\, Z_{1}^{l}\big)\right) \right|\\
 & = \Big|\hat{p}_{eq}-\sum_{i}\hat{p}_{f}(i)p_{\tilde{f}}(i)\Big|.
\end{align*}
Hence there exists constant $z_{1}^{l}$ such that 
\[
\left|\mathrm{Cov}\big(g(M),\,\tilde{g}(C)\,|\, Z_{1}^{l}=z_{1}^{l}\big) \right| \ge \Big|\hat{p}_{eq}-\sum_{i}\hat{p}_{f}(i)p_{\tilde{f}}(i)\Big|.
\]
We have 
\begin{align*}
\rho_{\mathrm{m}}(M;C) & \ge\frac{\big|\mathrm{Cov}\big(g(M),\,\tilde{g}(C)\,|\, Z_{1}^{l}=z_{1}^{l}\big)\big|}{\sqrt{\mathrm{Var}(g(M)\,|\, Z_{1}^{l}=z_{1}^{l})\mathrm{Var}(\tilde{g}(C)\,|\, Z_{1}^{l}=z_{1}^{l})}}\\
 & \ge\frac{\big|\mathrm{Cov}\big(g(M),\,\tilde{g}(C)\,|\, Z_{1}^{l}=z_{1}^{l}\big)\big|}{\sqrt{\E((g(M))^{2}\,|\, Z_{1}^{l}=z_{1}^{l})\E((\tilde{g}(C))^{2}\,|\, Z_{1}^{l}=z_{1}^{l})}}\\
 & \ge\frac{\big|\hat{p}_{eq}-\sum_{i}\hat{p}_{f}(i)p_{\tilde{f}}(i)\big|}{\sqrt{\chi^{2}+1}}.
\end{align*}
The result follows. 
\end{IEEEproof}

\smallskip{}

To obtain Theorem~\ref{thm:binary_guess}, take $\P$ to be the uniform distribution on $M$ and $\hat{\P}$ to be the actual distribution. The result follows from
%$\sum_{i}\hat{p}_{f}(i)p_{\tilde{f}}(i) \le \max_{i}\hat{p}_{f}(i)$
$\sum_{i} \hat{\P}\left\{ f(M)=i\right\} \cdot \P\{\tilde{f}(C)=i\} \le \max_{i}\hat{\P}\left\{ f(M)=i\right\}$
and
\[
\chi^{2}\left(p_{M}\Vert\mathrm{Unif}[1:2^{n}]\right)=2^{n-H_{2}(M)}-1.
\]

To show the result in Remark~\ref{rmk:ent1}, take $\tilde{F}$ to be a random variable in the probability space $\hat{\P}$ independent of $M$ with $\hat{\P}\{\tilde{F} = i\} = \P\{\tilde{f}(C)=i\}$, then the result follows from
\[
\hat{\P}\{ f(M)=\tilde{F}\} = \sum_{i} \hat{\P}\left\{ f(M)=i\right\} \cdot \P\{\tilde{f}(C)=i\}.
\]

Then we prove a slightly better result for one-bit functions $f(m)$.
Note that it is implied by the following more general result.
\begin{prop}
\label{prop:binary_guess_dist} Consider any two pmfs $p_{M}(m)$
and $\hat{p}_{M}(m)$ on $\Mc$, and a Markov kernel $p_{C|M}(c|m)$.
The two pmfs induce the joint probability measures \textnormal{$\P$} and \textnormal{$\hat{\P}$}
on $(M,C)$, respectively. Let $\rho_{\mathrm{m}}(M;C)$ be the maximal
correlation in \textnormal{$\P$}. For any one-bit functions $f:\Mc\to\left\{ 0,1\right\} $
and $\tilde{f}:\Cc\to\left\{ 0,1\right\} $, we have \textnormal{
\[
\hat{\P}\left\{ f(M)=\tilde{f}(C)\right\} -\frac{1}{2}\le\sqrt{\frac{1}{4}\rho_{\mathrm{m}}^{2}(M;C)\left(\chi^{2}\left(\hat{p}_{M}\Vert p_{M}\right)+1\right)+\left(1-\rho_{\mathrm{m}}^{2}(M;C)\right)\left(\hat{\P}\left\{ f(M)=0\right\} -\frac{1}{2}\right)^{2}}.
\]
} \end{prop}
\begin{IEEEproof}
All expectations, variances and covariances in this proof are in $\P$.
Let $g(m)=\left(-1\right)^{f(m)}\hat{p}_{M}(m)/p_{M}(m)$ and $\tilde{g}(c)=\left(-1\right)^{\tilde{f}(c)}$.
Write $\chi^{2}=\chi^{2}\left(\hat{p}_{M}\Vert p_{M}\right)$, $\hat{p}_{f}(i)=\hat{\P}\left\{ f(M)=i\right\} $,
$p_{\tilde{f}}(i)=\P\{\tilde{f}(C)=i\}$ and $\hat{p}_{e}=\hat{\P}\{f(M)\neq\tilde{f}(C)\}$.
It is straightforward to check that Proposition~\ref{prop:binary_guess_dist}
is true if $\hat{p}_{e}\ge\min\{\hat{p}_{f}(0),\hat{p}_{f}(1)\}$.
Hence we assume $\hat{p}_{e}<\min\{\hat{p}_{f}(0),\hat{p}_{f}(1)\}$.
Observe that 
\[
\E(g(M))=\hat{p}_{f}(0)-\hat{p}_{f}(1),
\]
\[
\E(\tilde{g}(C))=p_{\tilde{f}}(0)-p_{\tilde{f}}(1),
\]
\begin{align*}
\mathrm{Var}(g(M)) & =\sum_{m}\frac{\left(\hat{p}_{M}(m)\right)^{2}}{p_{M}(m)}-\left(\hat{p}_{f}(0)-\hat{p}_{f}(1)\right)^{2}\\
 & =(\chi^{2}+1)-\left(\hat{p}_{f}(0)-\hat{p}_{f}(1)\right)^{2},
\end{align*}
\[
\mathrm{Var}(\tilde{g}(C))=1-\left(p_{\tilde{f}}(0)-p_{\tilde{f}}(1)\right)^{2},
\]
\begin{align*}
\mathrm{Cov}\big(g(M),\,\tilde{g}(C)\big) & =\E\big(g(M)\tilde{g}(C)\big)-\E(g(M))\E(\tilde{g}(C))\\
 & =\sum_{m}p(m)g(m)\sum_{c}p(c|m)\tilde{g}(c)-\big(\hat{p}_{f}(0)-\hat{p}_{f}(1)\big)\big(p_{\tilde{f}}(0)-p_{\tilde{f}}(1)\big)\\
 & =1-2\hat{p}_{e}-\left(\hat{p}_{f}(0)-\hat{p}_{f}(1)\right)\left(p_{\tilde{f}}(0)-p_{\tilde{f}}(1)\right).
\end{align*}
We have 
\begin{align*}
\rho_{\mathrm{m}}^{2}(M;C) & \ge\frac{\left(\mathrm{Cov}\big(g(M),\tilde{g}(C)\big)\right)^{2}}{\mathrm{Var}\big(g(M)\big)\mathrm{Var}\big(\tilde{g}(C)\big)}\\
 & =\frac{\left(1-2\hat{p}_{e}-\big(\hat{p}_{f}(0)-\hat{p}_{f}(1)\big)\big(p_{\tilde{f}}(0)-p_{\tilde{f}}(1)\big)\right)^{2}}{\left((\chi^{2}+1)-\left(\hat{p}_{f}(0)-\hat{p}_{f}(1)\right)^{2}\right)\left(1-\left(p_{\tilde{f}}(0)-p_{\tilde{f}}(1)\right)^{2}\right)}\\
 & \ge\frac{\left(1-2\hat{p}_{e}\right)^{2}-\left(1-2\hat{p}_{f}(0)\right)^{2}}{(\chi^{2}+1)-\left(1-2\hat{p}_{f}(0)\right)^{2}},
\end{align*}
where the last inequality is due to $\hat{p}_{e}<\min\{\hat{p}_{f}(0),\hat{p}_{f}(1)\}$
and 
\[
\frac{(a-bx)^{2}}{1-x^{2}}=\left(\frac{a}{\sqrt{1-x^{2}}}-\frac{bx}{\sqrt{1-x^{2}}}\right)^{2}\ge a^{2}-b^{2}
\]
for any $a,b$ such that $|a|>|b|$ and $-1<x<1$. Hence, 
\begin{align*}
\left(1-2\hat{p}_{e}\right)^{2} & \le\rho_{\mathrm{m}}^{2}(M;C)\left((\chi^{2}+1)-\left(1-2\hat{p}_{f}(0)\right)^{2}\right)+\left(1-2\hat{p}_{f}(0)\right)^{2}\\
 & =\rho_{\mathrm{m}}^{2}(M;C)(\chi^{2}+1)+\left(1-\rho_{\mathrm{m}}^{2}(M;C)\right)\left(1-2\hat{p}_{f}(0)\right)^{2}.
\end{align*}
%\[
%\rho_\mathrm{m}^{2}(M;C)\left((\chi^2+1)\left(1-2\hat{p}_{f}(0)\right)^{2}\right)+\left(1-2\hat{p}_{f}(0)\ right)^{2}\ge\left(1-2\hat{p}_{e}\right)^{2},
%\]
The result follows. 
\end{IEEEproof}
\smallskip{}

To prove Theorem \ref{thm:binary_guess} for one-bit functions $f(m)$,
note that by Proposition \ref{prop:binary_guess_dist}, \begingroup
\allowdisplaybreaks 
\begin{align*}
\hat{\P}\left\{ f(M)=\tilde{f}(C)\right\} -\frac{1}{2} & \le\sqrt{\frac{1}{4}\rho_{\mathrm{m}}^{2}(M;C)\left(\chi^{2}\left(\hat{p}_{M}\Vert p_{M}\right)+1\right)+\left(1-\rho_{\mathrm{m}}^{2}(M;C)\right)\left(\hat{\P}\left\{ f(M)=0\right\} -\frac{1}{2}\right)^{2}}\\
 & \le\sqrt{\frac{1}{4}\rho_{\mathrm{m}}^{2}(M;C)\left(\chi^{2}\left(\hat{p}_{M}\Vert p_{M}\right)+1\right)+\left(\hat{\P}\left\{ f(M)=0\right\} -\frac{1}{2}\right)^{2}}\\
 & \le\sqrt{\frac{1}{4}\rho_{\mathrm{m}}^{2}(M;C)\left(\chi^{2}\left(\hat{p}_{M}\Vert p_{M}\right)+1\right)}+\sqrt{\left(\hat{\P}\left\{ f(M)=0\right\} -\frac{1}{2}\right)^{2}}\\
 & =\frac{1}{2}\rho_{\mathrm{m}}(M;C)\sqrt{\chi^{2}\left(\hat{p}_{M}\Vert p_{M}\right)+1}+\max\left\{ \hat{\P}\left\{ f(M)=0\right\} ,\,\hat{\P}\left\{ f(M)=1\right\} \right\} -\frac{1}{2}.
\end{align*}
\endgroup This completes the proof.

%------------------------------

\subsection{Proof of Theorem \ref{thm:converse}\label{subsec:Proofs-Thm-Converse}}

Here we assume the ciphertext length is arbitrary, and the encryption
function can be randomized, i.e., the ciphertext is $C=E(K,M,W)$
where $W\sim p(w)$ is the local randomness at the sender. We require
that $D(k,E(k,m,w))=m$ for any $k,m,w$. From Lemma \ref{lem:rho_chisq},
\begin{align*}
\rho^{2} & \ge\rho_{\mathrm{m}}^{2}(M;C)\\
 & \ge2^{-n}\chi^{2}\left(p(m,c)\,\Vert\, p(m)p(c)\right)\\
 & =2^{-n}\left(\sum_{m,c}\frac{\left(p(m,c)\right)^{2}}{p(m)p(c)}-1\right)\\
 & =\sum_{m,c}\frac{\left(p(m,c)\right)^{2}}{p(c)}-2^{-n}\\
 & =\sum_{c}\frac{\sum_{m}\left(p(m,c)\right)^{2}}{p(c)}-2^{-n}\\
 & \ge\sum_{c}\frac{\left|\left\{ m:\, p(m,c)>0\right\} \right|^{-1}\left(\sum_{m}p(m,c)\right)^{2}}{p(c)}-2^{-n}\\
 & =\E\left(\left|\left\{ m:\, p(m,C)>0\right\} \right|^{-1}\right)-2^{-n}\\
 & \ge\E\left(\left|\left\{ D(k,C):\, k\in[1:2^{s}]\right\} \right|^{-1}\right)-2^{-n}\\
 & \ge2^{-s}-2^{-n}.
\end{align*}
Hence, 
\begin{align*}
s & \ge\log\left(\frac{1}{\rho^{2}+2^{-n}}\right)\\
 & =2\log\frac{1}{\rho}-\log\left(1+\frac{1}{2^{n}\rho^{2}}\right).
\end{align*}
This completes the proof. %-------------------------------------------------------

\subsection{Proof of Theorem \ref{thm:achieve_expander}\label{subsec:Proofs-Thm-Expander}}

Theorem \ref{thm:achieve_expander} is a direct consequence of Lemma
\ref{lem:rho_norm}. Since $M,C\sim\U[1:2^{n}]$, we have 
\begin{align*}
\rho_{\mathrm{m}}(M;C) & =2^{n}\left\Vert P_{MC}-2^{-2n}\mathbf{1}_{2^{n}\times2^{n}}\right\Vert \\
 & =2^{n}\left\Vert \frac{1}{d\cdot2^{n}}A-2^{-2n}\mathbf{1}_{2^{n}\times2^{n}}\right\Vert \\
 & =\frac{1}{d}\left\Vert A-\frac{d}{2^{n}}\mathbf{1}_{2^{n}\times2^{n}}\right\Vert \\
 & =\frac{1}{d}|\lambda_{2}(A)|.
\end{align*}
Ramanujan graphs have second eigenvalue $|\lambda_{2}(A)|\le2\sqrt{d-1}$,
hence their maximal correlation is 
\begin{align*}
\rho_{\mathrm{m}}(M;C) & \le\frac{2\sqrt{d-1}}{d}\\
 & \le\frac{2}{\sqrt{d}}.
\end{align*}
As a result, if $\log d\ge2\log(1/\rho)+2$, we have $\rho_{\mathrm{m}}(M;C)\le\rho$.
%------------------------------

\subsection{Proofs of Theorem \ref{thm:rand_achieve} and Corollary~\ref{cor:rand_achieve_rhoonly}\label{subsec:Proofs-Thm-Rand}}

We first prove a lemma on the cascade of two ciphers with the same
message length $n$ but with possibly different key lengths $s_{1}$
and $s_{2}$, which yields a cipher with message length $n$ and key
length $s_{1}+s_{2}$.
\begin{lem}
\label{lem:cipher_combine} Let $(E_{1},D_{1})$ and $(E_{2},D_{2})$
be two ciphers with key lengths $s_{1}$ and $s_{2}$, respectively,
and the same message length $n$. Define the cascade of these two
ciphers to be the cipher $\mathcal{K}=[1:2^{s_{1}}]\times[1:2^{s_{2}}]$,
$\mathcal{M}=\mathcal{C}=[1:2^{n}]$,{
\[
E(k_{1},k_{2},m)=E_{2}\left(k_{2},E_{1}(k_{1},m)\right),\; D(k_{1},k_{2},m)=D_{1}\left(k_{1},D_{2}(k_{2},m)\right).
\]
} Then we have{
\[
\rho_{\mathrm{m}}\left(M;\, E(K_{1},K_{2},M)\right)\le\rho_{\mathrm{m}}\left(M;\, E_{1}(K_{1},M)\right)\cdot\rho_{\mathrm{m}}\left(M;\, E_{2}(K_{2},M)\right).
\]
} \end{lem}
\begin{IEEEproof}
Consider the following alternate characterization of maximal correlation
in \cite{renyi1959measures} 
\[
\rho_{\mathrm{m}}\left(X;\, Y\right)=\max_{f(x):\,\E(f(X))=0,\,\E(f^{2}(X))=1}\sqrt{\E\left(\left(\E\left(f(X)\,|\, Y\right)\right)^{2}\right)}.
\]
Let $M_{1}\sim\U[1:2^{n}]$, $M_{2}=E_{1}(K_{1},M_{1})$, $C=E_{2}\left(K_{2},M_{2}\right)$.
Note that for any $f,g:[1:2^{n}]\to\mathbb{R}$ with $\E(f(M_{1}))=\E(g(C))=0$,
$\E(f^{2}(M_{1}))=\E(g^{2}(C))=1$, by the alternate characterization,
\begin{align*}
\E\left(f(M_{1})g(C)\right) & =\E\left(\E\left(f(M_{1})\,|\, M_{2}\right)\cdot\E\left(g(C)\,|\, M_{2}\right)\right)\\
 & \le\sqrt{\E\left(\left(\E\left(f(M_{1})\,|\, M_{2}\right)\right)^{2}\right)\cdot\E\left(\left(\E\left(g(C)\,|\, M_{2}\right)\right)^{2}\right)}\\
 & \le\rho_{\mathrm{m}}\left(M_{1};M_{2}\right)\cdot\rho_{\mathrm{m}}\left(C;M_{2}\right).
\end{align*}
The result follows.
\end{IEEEproof}
Now consider the following result from~\cite{Friedman1991}. Let
$A_{1},\ldots,A_{d}\in\mathbb{R}^{N\times N}$ be i.i.d. random permutation
matrices uniformly distributed in the set of permutations of $\left\{ 1,\ldots,N\right\} $.
Then we have 
\begin{equation}
\E\left(\left|\lambda_{2}\left(\sum_{i=1}^{d}\left(A_{i}+A_{i}^{T}\right)\right)\right|\right)\le2\sqrt{2d-1}\left(1+\frac{\ln d}{\sqrt{2d}}+O\left(d^{-1/2}\right)\right)+O\left(\frac{d^{3/2}\ln\ln N}{\ln N}\right).\label{eq:sum_rand_perm}
\end{equation}

We use the above result to construct a cipher as follows. Generate
$d=2^{s-1}$ permutations on $[1:2^{n}]$, namely $\sigma_{1},\ldots,\sigma_{d}$,
i.i.d. uniformly at random. Let $\sigma_{i+d}=\sigma_{i}^{-1}$ for
$i=1,\ldots,d$. The cipher is defined as $\mathcal{K}=[1:2^{s}]$,
$\mathcal{M}=\mathcal{C}=[1:2^{n}]$, $E(k,m)=\sigma_{k}(m)$, $D(k,c)=\sigma_{k}^{-1}(c)$.
By Lemma \ref{lem:rho_norm}, 
\begin{align*}
\rho_{\mathrm{m}}(M;C) & =\left\Vert 2^{n}P_{MC}-\frac{1}{2^{n}}\mathbf{1}_{2^{n}\times2^{n}}\right\Vert \\
 & =\left\Vert \frac{1}{2d}\sum_{i=1}^{d}\left(A_{i}+A_{i}^{T}\right)-\frac{1}{2^{n}}\mathbf{1}_{2^{n}\times2^{n}}\right\Vert \\
 & =\frac{1}{2d}\left|\lambda_{2}\left(\sum_{i=1}^{d}\left(A_{i}+A_{i}^{T}\right)\right)\right|.
\end{align*}
Hence by (\ref{eq:sum_rand_perm}), there exist fixed $\sigma_{1},\ldots,\sigma_{d}$
and a constant $\eta>0$ (that does not depend on $s$ or $n$) such
that 
\begin{align}
\rho_{\mathrm{m}}(M;C) & \le\frac{1}{2d}\left(2\sqrt{2d-1}\left(1+\frac{\ln d}{\sqrt{2d}}+\eta d^{-1/2}\right)+\eta\cdot\frac{d^{3/2}\log n}{n}\right)\nonumber \\
 & \le\frac{2}{\sqrt{2d}}\left(1+\frac{\ln d}{\sqrt{2d}}+\eta\left(d^{-1/2}+\frac{d\log n}{n}\right)\right)\nonumber \\
 & \le\frac{4}{\sqrt{2d}}\nonumber \\
 & =2^{-s/2+2}\label{eq:rho_small_s}
\end{align}
if $d\ge16\eta^{2}$ and $n/\log n\ge4\eta d$, or equivalently, 
\begin{align}
2\log\eta+5 & \le s\le\log n-\log\log n-\log\eta-1.\label{eq:s_cond}
\end{align}
Note that this construction only works for very short key lengths.
We now provide a construction for general key length $s$ by the cascade
of several ciphers with short key lengths. Let 
\[
t=\left\lceil \frac{s}{\log n-\log\log n-\log\eta-2}\right\rceil ,\,\tilde{s}=\left\lfloor \frac{s}{t}\right\rfloor ,\, a=t\left(\left\lfloor \frac{s}{t}\right\rfloor +1\right)-s,\, b=s-t\left\lfloor \frac{s}{t}\right\rfloor ,
\]
then we have $a+b=t$ and $s=a\tilde{s}+b(\tilde{s}+1)$. Consider
the cascade of $a$ ciphers with key length $\tilde{s}$ and $b$
ciphers of key length $\tilde{s}+1$, which gives a cipher with key
length $s$. Let $s_{0}$ be an integer satisfying 
\[
s_{0}\ge\max\{4\log\eta+12\,,\,2^{20}\}
\]
and 
\[
\log s_{0}-\log\log s_{0}\ge5\log\eta+14.
\]
Consider any $s\ge s_{0}$. If $n\le s$, then perfect secrecy can
be achieved. Hence we assume $n>s\ge s_{0}$. To check the conditions
in \eqref{eq:s_cond} for $\tilde{s}$ and $\tilde{s}+1$, 
\begin{align*}
\tilde{s} & =\bigg\lfloor\frac{s}{\Big\lceil s\left(\log n-\log\log n-\log\eta-2\right)^{-1}\Big\rceil}\bigg\rfloor\\
 & \ge\frac{s}{s\left(\log n-\log\log n-\log\eta-2\right)^{-1}+1}-1\\
 & =\frac{1}{\left(\log n-\log\log n-\log\eta-2\right)^{-1}+s^{-1}}-1\\
 & \ge\frac{1}{\left(4\log\eta+12\right)^{-1}+\left(4\log\eta+12\right)^{-1}}-1\\
 & =2\log\eta+5.
\end{align*}
And also 
\begin{align*}
\tilde{s}+1 & =\bigg\lfloor\frac{s}{\Big\lceil s\left(\log n-\log\log n-\log\eta-2\right)^{-1}\Big\rceil}\bigg\rfloor+1\\
 & \le\log n-\log\log n-\log\eta-1.
\end{align*}
Hence by~\eqref{eq:rho_small_s} and Lemma \ref{lem:cipher_combine},
the maximal correlation of the resultant cipher is 
\begin{align*}
\rho_{\mathrm{m}}(M;C) & \le\left(2^{-\tilde{s}/2+2}\right)^{a}\left(2^{-(\tilde{s}+1)/2+2}\right)^{b}\\
 & =2^{-s/2+2t},
\end{align*}
where \begingroup \allowdisplaybreaks 
\begin{align*}
2t & =2\left\lceil \frac{s}{\log n-\log\log n-\log\eta-2}\right\rceil \\
 & \le\frac{2s}{\log n-\log\log n-\log\eta-2}+2\\
 & \le\frac{2s}{\log n-\log\log n-\left(\log n-\log\log n\right)/5}+2\\
 & =\frac{5s/2}{\log n-\log\log n}+2\\
 & \le\frac{4s}{\log n}+2,
\end{align*}
\endgroup where the last inequality is due to $\log n\ge\log s_{0}\ge14$.
Therefore, 
\[
2\log\frac{1}{\rho_{\mathrm{m}}(M;C)}\ge s\left(1-\frac{8}{\log n}\right)-4.
\]
Rearranging, we have 
\begin{align*}
\\
s & \le\left(2\log\frac{1}{\rho_{\mathrm{m}}(M;C)}+4\right)\left(1-\frac{8}{\log n}\right)^{-1}\\
 & \le\left(2\log\frac{1}{\rho_{\mathrm{m}}(M;C)}\right)\left(1+\frac{16}{\log n}\right)+16.
\end{align*}
Hence if 
\[
s\ge\left(2\log\frac{1}{\rho}\right)\left(1+\frac{\alpha}{\log n}\right)+\alpha.
\]
where $\alpha=\max\{16,s_{0}\}$, then $s\ge s_{0}$, and $\rho_{\mathrm{m}}(M;C)\le\rho$.
This completes the proof of Theorem~\ref{thm:rand_achieve}. \smallskip{}

Now we prove Corollary~\ref{cor:rand_achieve_rhoonly}. If 
\[
s\ge\left(2\log\frac{1}{\rho}\right)\left(1+\frac{3\alpha/2}{\log(\log(1/\rho)+1)}\right),
\]
then 
\begin{align*}
\log\frac{1}{\rho}+1 & \le\frac{s}{2}\left(1+\frac{3\alpha/2}{\log(\log(1/\rho)+1)}\right)^{-1}+1\\
 & \le\frac{s}{2}+1\\
 & \le s
\end{align*}
due to the assumption that $s\ge2$. Hence,

\begin{align*}
s & \ge\left(2\log\frac{1}{\rho}\right)\left(1+\frac{3\alpha/2}{\log(\log(1/\rho)+1)}\right)\\
 & =\left(2\log\frac{1}{\rho}\right)\left(1+\frac{\alpha}{\log(\log(1/\rho)+1)}\right)+\frac{\alpha\log(1/\rho)}{\log(\log(1/\rho)+1)}\\
 & \ge\left(2\log\frac{1}{\rho}\right)\left(1+\frac{\alpha}{\log(\log(1/\rho)+1)}\right)+\alpha\\
 & \ge\left(2\log\frac{1}{\rho}\right)\left(1+\frac{\alpha}{\log s}\right)+\alpha\\
 & \ge\left(2\log\frac{1}{\rho}\right)\left(1+\frac{\alpha}{\log n}\right)+\alpha,
\end{align*}
where the last step is due to the assumption that $n>s$. This complete
the proof of Corollary~\ref{cor:rand_achieve_rhoonly}.

%-------------------------

\subsection{Proof of Theorem \ref{thm:achieve_binaryadd}\label{subsec:Proofs-Thm-Binaryadd}}

We first compute the maximal correlation of a binary additive stream
cipher. The following proposition follows by Fourier analysis of Boolean functions \cite{o2014analysis}; see~\cite{Calmon2014inertia}.
\begin{prop}
A binary additive stream cipher has a maximal correlation{
\[
\rho_{\mathrm{m}}(M;C)=\max_{v\in\left\{ 0,1\right\} ^{n}\backslash0^{n}}\left|\frac{1}{\left|\Kc\right|}\sum_{k\in\Kc}\left(-1\right)^{\sum_{l=0}^{n-1}v_{l}G_{l}(k)}\right|.
\]
} \end{prop}
We now proceed to prove Theorem \ref{thm:achieve_binaryadd}. %It
%is equivalent to showing that when we generate $\left(G(k)\right)_{i}$
%following $\mathrm{Bern}(1/2)$ i.i.d. across $k$ and $i$, then
%the probability of giving a binary additive stream cipher achieving
%$\rho$-maximal correlation secrecy is at least
%\[
%1-\exp\left(\left(\ln2\right)\left(n+1\right)-\frac{1}{2}\left|\Kc\right|\rho^{2}\right).
%\]
%Then we show that $\rho$-maximal correlation secrecy is achieved
%by most binary additive stream ciphers as long as $n\lesssim\rho^{2}\left|\Kc\right|$,
%in the sense that among the space of possible $G(k)$, most of them
%gives a binary additive stream cipher attaining $\rho$-maximal correlation
%secrecy.
%Assume we generate $\left(G(k)\right)_{i}$ following $\mathrm{Bern}(1/2)$
%i.i.d. across $k$ and $i$, then the maximal correlation of the randomly
%generated binary additive stream cipher satisfies 
%\begin{align*}
%\P\left\{ \rho_{\mathrm{Enc}}\ge\rho\right\}  & \le  2^{\left(n+1\right)-\left|\Kc\right|\cdot D_{\mathrm{KL}}\left(\left.\frac{1+\rho}{2}\right\Vert \frac{1}{2}\right)}.
%\end{align*}
Assume we generate $G_{i}(k)$ i.i.d. $\mathrm{Bern}(1/2)$ across
$k$ and $i$. For each fixed $v\neq0^{n}$, consider 
\[
\frac{1}{\left|\Kc\right|}\sum_{k\in\Kc}\left(-1\right)^{\sum_{l=0}^{n-1}v_{l}G_{l}(k)}.
\]
The terms $\left(-1\right)^{\sum_{l=0}^{n-1}v_{l}G_{l}(k)}$ are i.i.d.
Rademacher. By the Chernoff bound, 
\[
\P\left\{ \left|\frac{1}{\left|\Kc\right|}\sum_{k\in\Kc}\left(-1\right)^{\sum_{l=0}^{n-1}v_{l}G_{l}(k)}\right|\ge\rho\right\} \le2^{1-\left|\Kc\right|\cdot D_{\mathrm{KL}}\left(\left.\frac{1+\rho}{2}\right\Vert \frac{1}{2}\right)}.
\]
By the union bound on all possible $v\in\{0,1\}^{n}\backslash0^{n}$
and observing that $\left(\ln2\right)D_{\mathrm{KL}}\left(\left.\frac{1+\rho}{2}\right\Vert \frac{1}{2}\right)>\rho^{2}/2$
for $\rho>0$, 
\begin{align*}
\mathbb{P}\left\{ \rho_{\mathrm{m}}(M;\, C)\ge\rho\right\}  & \le2^{\left(n+1\right)-\left|\mathcal{K}\right|\cdot D_{\mathrm{KL}}\left(\left.\frac{1+\rho}{2}\right\Vert \frac{1}{2}\right)}\\
 & <2^{\left(n+1\right)-\left|\mathcal{K}\right|\rho^{2}/\left(2\ln2\right)}.
\end{align*}
Hence if 
\[
s\ge2\log\frac{1}{\rho}+\log n+\log\left(1+\frac{1}{n}\log\frac{1}{\e}\right)+2,
\]
then 
\begin{align*}
2^{s} & \ge4\rho^{-2}n\left(1+\frac{1}{n}\log\frac{1}{\e}\right)\\
 & \ge4\left(\ln2\right)\rho^{-2}\left(n+\log\frac{1}{\e}\right).
\end{align*}
Therefore, 
\begin{align*}
\mathbb{P}\left\{ \rho_{\mathrm{m}}(M;\, C)\ge\rho\right\}  & <2^{\left(n+1\right)-2^{s}\rho^{2}/\left(2\ln2\right)}\\
 & \le2^{\left(n+1\right)-2\left(n-\log\e\right)}\\
 & \le2^{1-n+\log\e}\\
 & \le\e.
\end{align*}
This completes the proof of Theorem \ref{thm:achieve_binaryadd}.
%------------------------------

%-----------------------------------

 \bibliographystyle{IEEEtran}
\bibliography{nit,ref}

\end{document}